\documentclass[conference]{IEEEtran}
\IEEEoverridecommandlockouts
\usepackage{amsmath}
\usepackage{balance}
\usepackage{cite}
\usepackage{acronym}
\usepackage{subfigure}
\usepackage{hyperref}
\usepackage[usenames]{color}

\usepackage[pdftex]{graphicx}

\acrodef{BRN}[BRN]{barrage relay network}
\acrodef{CBR}[CBR]{controlled barrage region}
\acrodef{OLA}[OLA]{opportunistic large array}
\acrodef{MANET}[MANET]{mobile ad hoc network}
\acrodef{TDMA}[TDMA]{time division multiple access}
\acrodef{GPS}[GPS]{Global Positioning System}
\acrodef{RTS}[RTS]{request-to-send}
\acrodef{CTS}[CTS]{clear-to-send}
\acrodef{SNR}[SNR]{signal-to-noise ratio}
\acrodef{SINR}[SINR]{signal-to-interference-plus-noise ratio}
\acrodef{CCI}[CCI]{co-channel interference}
\acrodef{TC}[TC]{transport capacity}
\acrodef{MLSE}[MLSE]{maximal-likelihood sequence estimation}
\acrodef{AWGN}[AWGN]{additive white Gaussian noise}
\acrodef{ASE}[ASE]{area spectral efficiency}
\acrodef{USSOCOM}[USSOCOM]{United States Special Operations Command}
\acrodef{IP}[IP]{Internet protocol}
\acrodef{PLI}[PLI]{position location information}

\begin{document}
\abovedisplayskip=0.15pt
\belowdisplayskip=0.15pt
\title{
Controlled Barrage Regions: \\
Stochastic Modeling, Analysis, and Optimization
 \author{Salvatore Talarico\IEEEauthorrefmark{1}, Matthew C. Valenti\IEEEauthorrefmark{1}, and Thomas R. Halford\IEEEauthorrefmark{3} \\
\IEEEauthorblockA{
\IEEEauthorrefmark{1}West Virginia University, Morgantown, WV, USA.\\
\IEEEauthorrefmark{2} WPL, Inc., Manhattan Beach, CA, USA.
 }
 }
 }
\date{}
\maketitle

\vspace{0.00cm}

\pagestyle{plain}

\begin{abstract}
A \acf{BRN} is a broadcast-oriented ad hoc network involving autonomous cooperative communication, a slotted time-division frame format, and a coarse slot-level synchronization.  While inherently a broadcast protocol, \acp{BRN} can support unicast transmission by superimposing a plurality of \acfp{CBR} onto the network.  Within each \acp{CBR}, a new packet is injected by the unicast source during the first time slot of each new radio frame. When a \acp{CBR} is sufficiently long that a packet might not be able to reach the other end within a radio frame, multiple packets can be active at the same time via spatial pipelining, resulting in interference within the \acp{CBR}. In this paper, the dynamics of packet transmission within a \acp{CBR} is described as a Markov process, and the outage probability of each link within the \acp{CBR} is evaluated in closed form, thereby accounting for fading and co-channel interference. In order to account for the linkage between simultaneous active packets and their temporal correlation, a Viterbi-like algorithm is used. Using this accurate analytical framework, a line network is optimized, which identifies the code rate, the number of relays, and the length of a radio frame that maximizes the transport capacity.\\
\end{abstract}

\vspace{-0.30cm}

\section{Introduction}
\vspace{-0.05cm}
In the past decade, significant research effort has been devoted to the topic of {\em cooperative communications} for wireless networks, and the corresponding benefits have been widely noted in the literature \cite{scaglione2006,stankovic2006,Sendonaris2003,Sendonaris2003b,Laneman2004,scaglione2003,chugg:2006}.
The use of Markov chains and their properties have also been extensively used to characterize and model the dynamics of such networks. However, most of the work in the literature (e.g.,\cite{marchenko2014,Hassan:2011}) neglects the effect of co-channel interference that is caused by multiple packets sharing the same radio resources or makes simplified assumptions (e.g.,\cite{jung:2013,Jung:2014}) in order to facilitate the analysis.
Furthermore, the outage probability of each link is usually computed by conditioning over a set of channel realizations, and then the average performance of the network is evaluated by taking the average over the fading, once the end-to-end outage probability is obtained. While conceptually straightforward, this approach is very computationally intense, and becomes computationally unfeasible as the size of the network (i.e., number of relays) increases.

The focus of this paper is a class of cooperative networks known as {\em \acfp{BRN}} \cite{halford:2010a}, whose waveform is going to be used as the \ac{MANET} solution for the next generation of \ac{USSOCOM} handheld radios to reliably deliver in harsh radio frequency environments cellular quality voice, high throughput video, \ac{IP} data, and  \ac{PLI} simultaneously in a single network. \acp{BRN} employ an efficient time-slotted flooding protocol wherein packets ripple out from sources in pipelined spatial waves. In a \ac{BRN}, simultaneous transmissions of the same packet are not suppressed, but rather exploited for the resulting diversity gains. The spatial extent of unicast and multicast transmissions can be contained in \acp{BRN} via {\em \acfp{CBR}}. Briefly, a \ac{CBR} is established by identifying a ring of buffer nodes around a portion of the network containing a source and its destination(s). Within the \ac{CBR}, the barrage-flooding primitive is used to transport data. The buffers suppress their relay function, and spatial pipelining is enabled, thereby enabling multiple \acp{CBR} and multiple packets within a \ac{CBR} to be active at the same time, causing respectively inter-\ac{CBR} and intra-\ac{CBR} interference.

\acp{BRN} resemble the \acp{OLA} introduced by Scaglione and Hong in \cite{scaglione2003}. Indeed, analogs to \acp{CBR} in \acp{OLA} have been proposed \cite{ingram:2009}. Both \acp{OLA} and \acp{CBR} exploit the diversity that can be obtained when multiple nodes transmit identical packets. \ac{BRN} can be distinguished from early \acp{OLA} descriptions by the time synchronization and receiver signal processing employed in \ac{BRN}, which enables packets to be longer and data rates to be larger than the inverse of the maximum relative delay spread between cooperating transmitters. The most important difference between more recent \ac{OLA} descriptions (e.g., \cite{ingram:2010}) and \ac{BRN} is the latter's use of \textit{autonomous} cooperative communications \cite{chugg:2006} as opposed to distributed space-time coding. This feature makes \acp{CBR} more suited for use in highly dynamic, tactical environments.

\vspace{-0.20cm}

\subsection{Contribution and Outline}

The main contribution of this paper is to model, analyze, and optimize unicast transmission in a \ac{BRN}, by accurately and efficiently tracking the dynamics of packet transmission within a \ac{CBR}.  Compared to related work in the literature, the paper accurately accounts for the co-channel interference, yet does so with minimal computational effort.  The key is to use the properties of a Markov process to track the dynamics of packet transmission within a \ac{CBR}, and the resulting formulation results in a {\em Viterbi-like} algorithm for performance evaluation.

More specifically, the time dynamics of a packet transmission within a \ac{CBR} is stochastically modeled using a Markov chain, and the transition probabilities are evaluated for a given network topology through a closed-form expression for the outage probability of each cooperative transmission \cite{Talarico2013}. The expression accounts for path loss, Rayleigh fading, thermal noise, and co-channel interference from multiple packets concurrently propagating within the network. By leveraging the properties of Markov chains and utilizing the closed-form expressions for transmission probabilities (which are already averaged over the fading), the average end-to-end outage probability and throughput is evaluated in a direct manner, greatly reducing the computational effort required for its evaluation.

While the effect of inter-\ac{CBR} interference is studied in \cite{talarico:2014}, here the effect of intra-\ac{CBR} interference, which occurs when a new packet is injected into a \ac{CBR} even before the previous packet reaches the end of the \ac{CBR}, is carefully and accurately considered. In order to capture the temporal correlation among packets and the reciprocal effect that interfering packets cause to each other, a Viterbi-like algorithm is used, which iteratively updates the transition probabilities of the Markov process on a frame-by-frame basis. Having established an analytical framework that can obtain exact expressions for the throughput of a given network configuration, this paper proceeds to optimize the network with respect to the {\em \ac{TC}}, which is a measure of forward progress.  In particular,  the network is optimized for several different scenarios  by selecting the length of a radio frame, the code rate per transmission, and the number of relays involved in the communication that maximizes  the \ac{TC}.

The remainder of the paper is organized as follows. Sections \ref{Section:BRN} and \ref{Section:CBR} overview \acp{BRN} and \acp{CBR}, respectively. Section \ref{Section:SystemModel} presents a system model and leverages a new expression \cite{Talarico2013}  for  the  conditional  outage  probability. Section \ref{Section:MarkovChain} describes how to model the dynamics of a packet within a \ac{CBR} using a Markov chain and its properties. Section \ref{Section:IterativeMethod} presents an iterative method that accounts for the interdependence between simultaneous active packets and their temporal correlation.
Section \ref{Section:Optimization} provides a framework for optimizing the network.  Finally, the paper concludes in Section \ref{Section:Conclusion}.

\section{Barrage Relay Network} \label{Section:BRN}

\acp{BRN} are a class of \acp{MANET} \cite{halford:2010a,halford:2010b}, which are based on an autonomous cooperative-communication scheme, that shares radio resources by using a combination of space-division and time-division.  Time-division is achieved by using a common frame comprised of $F$ slots, which are coarsely synchronized.  In the absence of GPS, the coarse slot-level synchronization can be achieved by using a distributed network-timing protocol.  In contrast with common conventional approaches for cooperative communications, which typically employ distributed space-time coding, relay selection, or distributed beamforming, in a \ac{BRN} each transmitting  node pseudo-randomly  dithers  its  carrier phase  in an independent manner so  that  the  superposition  of  these  signals will  induce a  time-varying  fading characteristic  at  a  receiving  node,  as detailed in \cite{chugg:2006}.

\acp{BRN} do not require any inter-node communication, and by changing the phase dither within a packet, destructive interference across an entire packet can be avoided. A forward error-correction code with long block lengths can then  be  used  to  extract  the  time  diversity provided by this induced time-varying fading channel, which is captured by \ac{MLSE} equalization and iterative detection.
The autonomous cooperation ensures that the concurrent transmission of identical packets from different nodes does not result in a collision or destructive superposition for all the nodes within a given range, but instead results in a form of cooperative diversity that increases the probability of successful reception. Furthermore, the protocol minimizes the overhead required for cooperation and enables multiple nodes to participate in the transmission without any knowledge regarding which other nodes are transmitting.

The \ac{BRN} flooding primitive works as follows. During the first slot of a frame, the source broadcasts a new packet.  During the second slot, any node that received and successfully decoded the initial transmission will rebroadcast it. If more than one node concurrently transmits the message, then the superposition of their signals is received, and the diverse signal components are effectively maximal-ratio combined by the \ac{MLSE} equalizer at each receiver. During each subsequent slot, all messages that were successfully decoded during the previous slot are rebroadcast. The process repeats until either the destination is reached, no receiver successfully decodes a transmission, or a maximum number of transmissions is reached.   Packets thus propagate outward from the source via a \emph{decode-and-forward} approach.  To prevent relay transmissions from propagating back towards the source, each node relays a given packet only once.

\begin{figure*}[ht]
\centering
\subfigure[Example of a \ac{CBR}. ]{\includegraphics[width=7 cm]{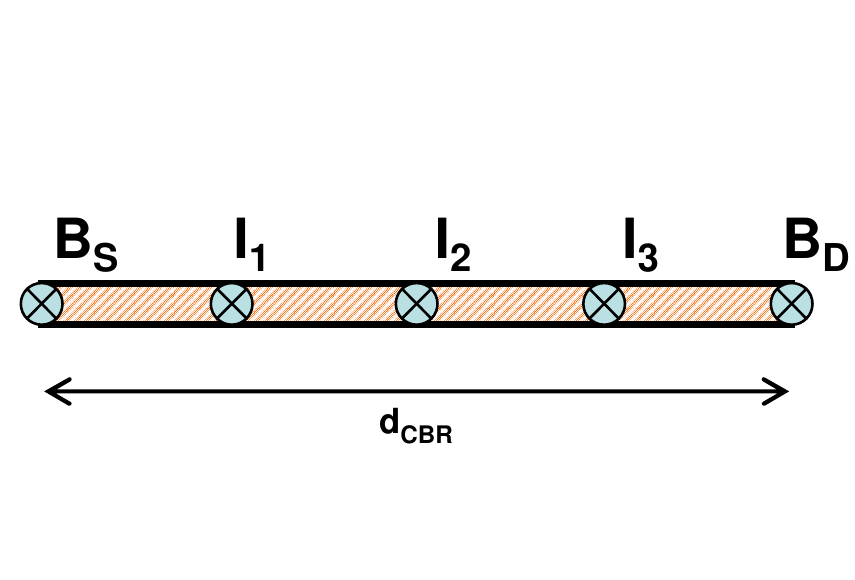}}
\subfigure[Time representation.]{\includegraphics[width=9cm]{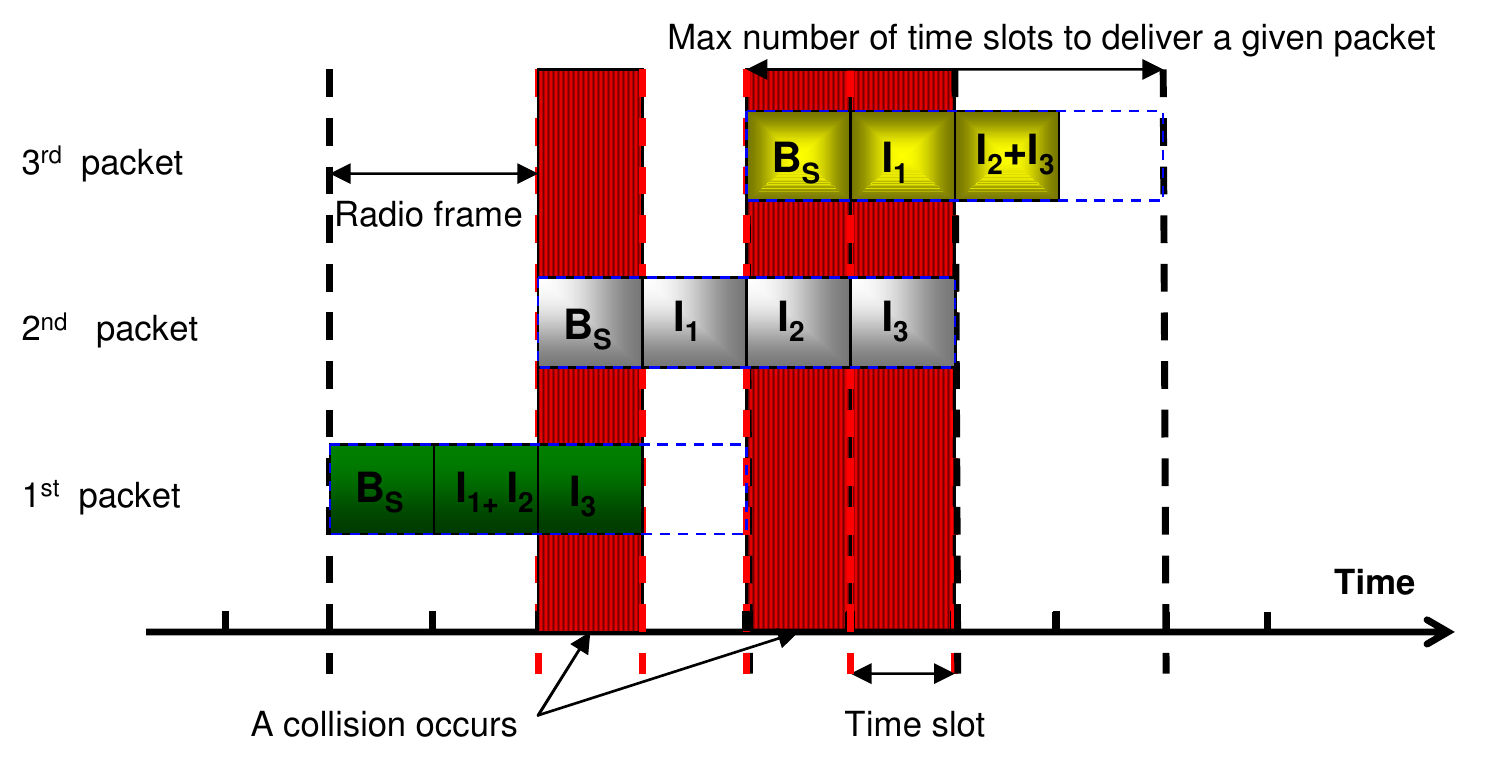}}
\vspace{-0.20cm}
\caption{Illustration of a typical \ac{CBR}, composed of three interior nodes ($N=3$), and an example of the possible dynamics in time of three consecutive packets within the network. The sub-figure on the right emphasizes the collisions that occur between packets, when they are simultaneously active. Each horizontal colored block indicates a packet, and for each time slot it is indicated within the block which node or nodes are transmitting. The red vertical block are used to indicate those slots during which a collision takes place.}
 \label{Example_CBR}
 \vspace{-0.6cm}
\end{figure*}

\section{Controlled Barrage Regions} \label{Section:CBR}

The time-slotted protocol enables spatial reuse by pipelining packets into a \ac{CBR} for transmission every $F$ slots and allows a \ac{BRN} to support multiple and contemporaneous  localized  unicast  streams, which can enhance its performance. A \ac{CBR} is the building block for such  spatially  separated  unicast  transmissions. \acp{CBR} can be established by specifying a set of \emph{buffer} nodes around a set of cooperating \emph{interior} nodes. Each buffer node acts as the source for transmissions into a \ac{CBR} and the destination for transmissions emanating from within the \ac{CBR}. The  relay  function of buffer nodes within a \ac{CBR} is suppressed so that external packets do not propagate into a given \ac{CBR}, nor do internal packets propagate to the rest of the network. In  this  way,  multiple unicast transmissions may be established in different portions of the network, enabling \emph{spatial reuse}. The buffer and interior nodes corresponding to a given source-destination  pair can be specified via the broadcast of \ac{RTS} / \ac{CTS} packets as described in \cite{halford:2010b,halford:2011}. This mechanism  for  unicast transport is more  robust  than  both  traditional unicast routing, and non-cooperative multipath variants.

Let $N$ be the number of the interior nodes (i.e., relays) of an established \ac{CBR}. A packet may be broadcast by the source buffer node of the \ac{CBR} during only the first time slot of a frame.  In ideal channel conditions, it is possible that the destination buffer node receives the initial source transmission in the first slot, though more generally, additional transmissions from the interior nodes may be required for successful reception. In the worst case scenario, a packet injected in a given \ac{CBR} can be received by its destination buffer node in a maximum of $N+1$ time slots (i.e., when there is no cooperation and all the interior relay nodes are used). In a typical scenario, $N>F+1$, and, due to spatial pipelining, multiple packets can be simultaneously active within a \ac{CBR}. Since these packets share the same radio resources they interfere with each other.

An illustration of this scenario is shown in Fig. \ref{Example_CBR}. Fig. \ref{Example_CBR}(a) shows an example for a linear \ac{CBR} of length $d_\text{CBR}$, which is composed of five nodes: three interior nodes, $I_1$, $I_2$, and $I_3$, which act as relays, and the two other nodes, which are buffers and act as the source buffer node $B_S$ and destination buffer node $B_D$  for the transmissions within the \ac{CBR}, respectively. In this example, assume that a radio frame is composed of $F=2$ time slots, which is the interval of time between the first transmission of a given packet and the first transmission of the next packet, since each new packet is transmitted during the first time slot of each new radio frame. In this scenario, in the worst case scenario a packet might be active for at most four consecutive time slots, which will correspond to the case when four hops are used to reach the destination buffer node, and no cooperation is used among nodes.

Fig. \ref{Example_CBR} (b) provides an example of how three consecutive packets might collide to each other. In this sub-figure, each horizontal colored rectangle indicates an active packet. Each horizontal rectangle is partitioned in a plurality of smaller blocks, which indicate the time slot during which the given packet is still active. For each time slot, it is indicated within a block which node or nodes are transmitting: for instance, the first packet is initially broadcast by the source, then during the second time slot $I_1$ and $I_2$ jointly rebroadcast it, which is finally diversity combined at $I_3$ and rebroadcast for the last time during the third time slot. The red vertical blocks are used to indicate those slots during which a collision takes place, while the dashed blue blocks that surround the packets indicate the maximum number of slots that might take the packet to be successfully received by $B_D$. In this example, the first and the second packet collide between each other, since the first packet is still active and is broadcast by $I_3$, when the second packet is injected for the first time within the \ac{CBR}. The second packet collides with the third packet for the duration of the entire first radio frame during which the third packet is first broadcast, since the second packets requires four hops to be delivered to $B_D$.

\section{Network Model} \label{Section:SystemModel}

Consider a \ac{BRN} composed of $K$ \acp{CBR} and $M$ mobile radios or \emph{nodes} $\mathcal X = \{X_1, ..., X_M\}$. Let $X_i$ represent both the $i^{th}$ node and its location. In the following, $X_i$ is typically used to denote a transmitting node and $X_j$ is typically used to describe a receiving node. During the $t^{th}$ time slot of a frame, $t\in\{1,...,F\}$, node $X_i$ transmits the $n^{th}$ packet with probability $p_{i,n}^{(t)}$.  Since the source of a given \ac{CBR} may transmit a new packet only during the first slot of a frame, for this node $p_{i,n}^{(t)}=0$ for $t \in \{2, ..., F\}$. The value of the transmit probability $p_{i,n}^{(t)}$ for each of the nodes depends on the dynamics of the protocol, and it can be evaluated as described in the following section.

Let $\mathcal X_{j,n}^{(t)} \subset \mathcal X$ be the set of cooperating or {\em barraging} nodes that transmit an identical copy of the $n^{th}$ packet to $X_j$ during the $t^{th}$ time slot and $\mathcal G_{j,n}^{(t)}$ the set of the indexes of the nodes in $\mathcal X_{j,n}^{(t)}$.
As described above, based on the dynamics of the protocol, nodes within the same \ac{CBR} as $X_j$, but not in $\mathcal X_{j,n}^{(t)}$ can transmit and broadcast a different packet, causing intra-\ac{CBR} interference. However, here the effect of the other active nodes in other \acp{CBR} is neglected, but it can be handled by concatenating the approach described in the following with the one described in \cite{talarico:2014}.

The power of the transmitter $X_i$ at the receiver $X_j$ during time slot $t$, when the $n^{th}$ packet is transmitted, is
\begin{eqnarray}
  \rho_{i,j,n}^{(t)}
  & = &
  P_i g_{i,j,n}^{(t)} f( d_{i,j} ) \label{eqn:power}
\end{eqnarray}
where $P_i$ is the averaged transmit power for $X_i$ in absence of fading at a reference  distance $d_{0}$, $g_{i,j,n}^{(t)}$ is the power gain due to fading for the link between node $X_j$ and node $X_i$, when the $n^{th}$ packet is transmitted, $d_{i,j} = ||X_i - X_j||$ is the distance from $X_j$ to $X_i$, and $f( \cdot )$ is a path-loss function.
The $\{ g_{i,j,n}^{(t)} \}$ are independent and exponentially distributed with unit mean, corresponding to Rayleigh fading. It is assumed that the \{$g_{i,j,n}^{(t)}\}$ remain fixed for the duration of a time slot, but vary independently from slot to slot. By assuming a distance-dependent path-loss function, for $d\geq d_{0}$, the attenuation power law is given as
\begin{eqnarray}
   f \left( d \right)
   & = &
   \left( \frac{d}{d_0} \right)^{-\alpha} \label{eqn:pathloss}
\end{eqnarray}
where $\alpha > 2$ is the path loss exponent, and $d_0$ is sufficiently large that the signals are in the far field.

When analyzing diversity combining, a commonly accepted approach is to assume that
the interference is fully-correlated among the branches (cf., \cite{Aalo2000}). While this condition is not always met, this makes the analysis tractable and it leads to a worse-case scenario. In particular, an upper bound on outage probability can be obtained, which happens to be tight \cite{Tanbourgi2013}. Under this assumption and by using (\ref{eqn:power}) and (\ref{eqn:pathloss}), the instantaneous \ac{SINR} at mobile $X_j$ during slot $t$ when the $n^{th}$ packet is transmitted is
\begin{eqnarray}
   \gamma_{j,n}^{(t)}
   & = &
   \frac{\displaystyle \sum_{k \in \mathcal G_{j,n}^{(t)}} g_{k,j,n}^{(t)} \Omega_{k,j}  }{ \displaystyle \Gamma^{-1} + \sum_{i \not\in \mathcal G_{j,n}^{(t)}, \forall z  \neq n} I_{i,z}^{(t)} g_{i,j,z}^{(t)} \Omega_{i,j} }
   \label{Equation:SINR2}
\end{eqnarray}
where $\Gamma = d_0^\alpha P/\mathcal{N}$ is the \ac{SNR} of a unit-distance transmission when fading is absent, $\Omega_{i,j} = d_{i,j}^{-\alpha}$ is the relative path gain, $I_{i,z}^{(t)}$ is a Bernoulli random variable indicating the $X_i$ is a source of interference during slot $t$, since it transmits the $z^{th}$ packet while the $n^{th}$ packet is still propagating, and $P[I_{i,z}^{(t)}=1]=p_{i,z}^{(t)}$.

\subsection{Outage probability}

Let $\beta$ denote the minimum \ac{SINR} required by $X_j$ for reliable reception and $\boldsymbol{\Omega }_j=\{\Omega_{1,j},...,\Omega _{M,j}\}$ represent the set of relative path gains from all $\{ X_i\}$ to $X_j$.  An \emph{outage} occurs when the \ac{SINR} falls below $\beta$. Conditioning on the path gains $\boldsymbol{\Omega }_j$ and the set of barraging nodes $\mathcal X_{j,n}^{(t)}$, the outage probability of mobile $X_j$ during slot $t$, when the $n^{th}$ packet is transmitted, is
\begin{eqnarray}
   \epsilon_{j,n}^{(t)}
   & = &
   P \left[ \gamma_{j,n}^{(t)} \leq \beta \Big| \boldsymbol \Omega_j, \mathcal X_{j,n}^{(t)} \right].
   \label{Equation:Outage1}
   \vspace{0.3cm}
\end{eqnarray}
Because it is conditioned on $\boldsymbol{\Omega }_j$, the outage probability depends on the particular network realization, which has dynamics over timescales that are much slower than the fading. From \cite{Talarico2013}, under the assumption that none of the baraging transmitters is collocated in the same position, and using the methodology of \cite{valenti:2014}, the outage probability in Rayleigh fading  is
\begin{eqnarray}
 \epsilon_{j,n}^{(t)} \hspace{-0.25cm}&=& \hspace{-0.25cm}
1 - \hspace{-0.1cm}\sum_{k \in \mathcal G_{j,n}^{(t)}} \hspace{-0.1cm} \exp\left(-\frac{\beta}{\Omega_{k,j} \Gamma }\right) \prod_{s \in \mathcal G_{j,n}^{(t)},s\neq k} \frac{\Omega_{k,j}}{\Omega_{k,j}-\Omega_{s,j}} \nonumber
\end{eqnarray}
\begin{eqnarray}
  \hspace{-0.25cm} \hspace{-0.25cm}
\times \prod_{ i \notin \mathcal G_{j,n}^{(t)}, \forall z  \neq n} \frac{ \Omega_{k,j}+\beta\left( 1-p_{i,z}^{(t)} \right) \Omega_{i,j}}{\Omega_{k,j}+ \beta \Omega_{i,j}}. \label{eqn_final_case1_Naka2}
\end{eqnarray}
The outage probability is conditioned on the node locations (represented by the $\{\Omega_{i,j}\}$) and by the particular set of barraging transmitters (represented by $\mathcal X_{j,n}^{(t)}$). 

\section{Modeling of the Dynamics of a Packet}\label{Section:MarkovChain}

The dynamics in time of a given packet within a \ac{CBR} can be described as an \emph{absorbing} Markov process. Define $\boldsymbol{s} = \{ s_1, s_2, \dots , s_{\varrho} \}$, which is the {\em state space} of the process composed of $\varrho$ {\em Markov} states $s_i$.  An absorbing Markov process is characterized by $\mathcal{\tau}$ {\em transient states} and $r$ {\em absorbing states}.  A state $s_i$ of a Markov chain is called \emph{absorbing} if once in that state, it is impossible to leave it, while a state that is not absorbing is called a \emph{transient} state.

Consider a single packet being injected within a \ac{CBR}, which is composed of a source (denoted $B_S$), a destination ($B_D$) and $N$ interior nodes or relays ($\{I_1, ...,I_N\}$). At the boundary between any two time slots, each node can be in one of the following states, which we refer to as the {\em node state}:
\begin{itemize}
\item Node state $0_n$: The node has not yet successfully decoded the $n^{th}$ packet.
\item Node state $1_n$: It has just decoded the $n^{th}$  packet received during previous slot, and it will transmit on next slot.
\item Node state $2_n$: It has decoded the $n^{th}$  packet in an earlier slot and it will no longer transmit or receive that packet, but it might be employed for the reception and transmission of a different packet.
\end{itemize}

\begin{figure}[t]
\centering
\includegraphics[width=9cm]{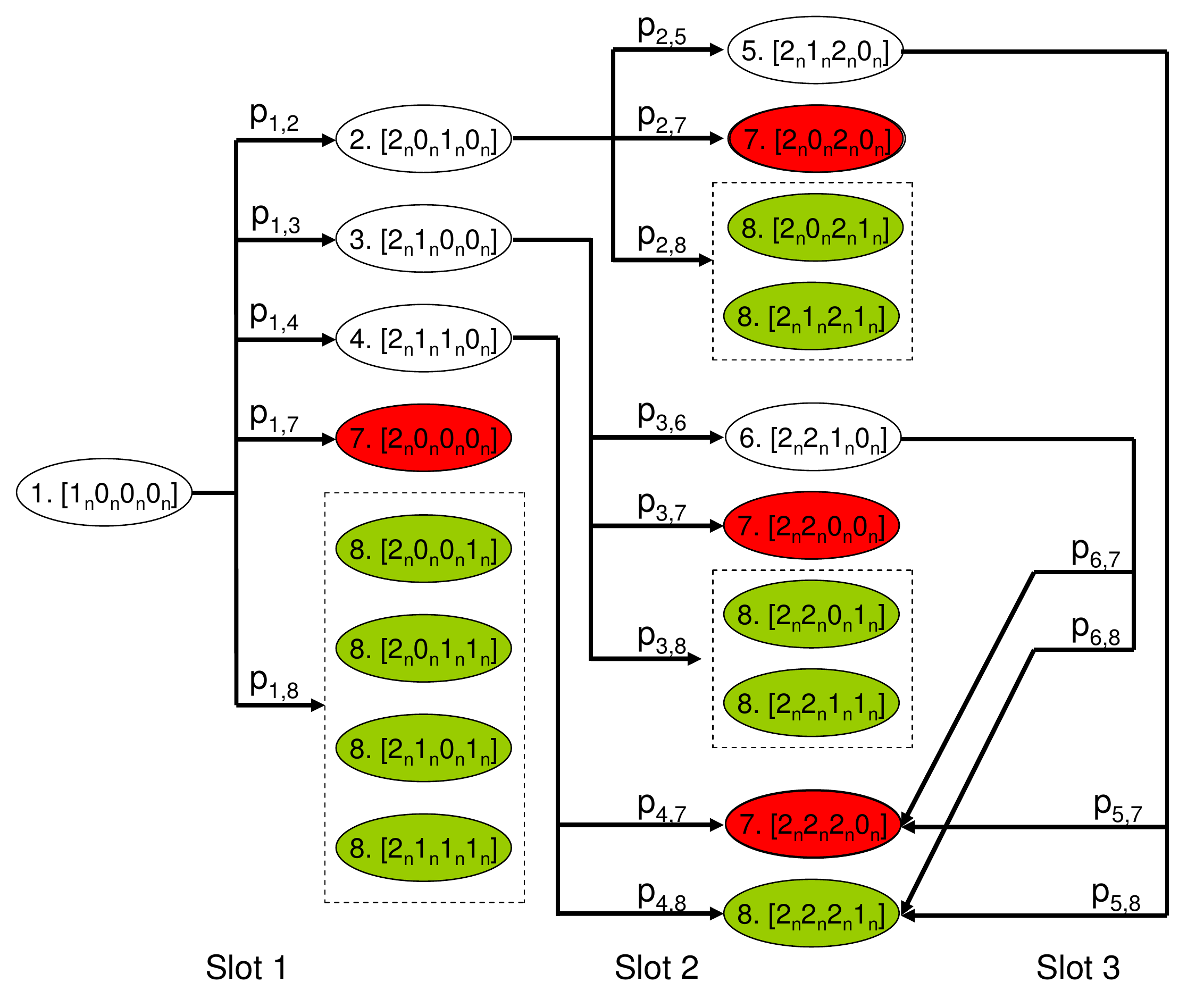}
\caption{Markov chain for a single packet that propagates within a \ac{CBR} composed of four nodes ($N=2$).  Transient states are in white, while the \ac{CBR} success absorbing state is in green and the \ac{CBR} outage absorbing state is in red.  Each of the two absorbing states is the union of several \ac{CBR} states.}
 \label{Markov_Chain}
 \vspace{-0.75cm}
\end{figure}

The state of a \ac{CBR} is the concatenation of the states of the individual nodes within the \ac{CBR}, which indicate the packet to which we are referring. Hereafter, we refer to this as the {\em \ac{CBR} state}.  The \ac{CBR} state can be compactly represented by a vector of the form $\left[ B_S, I_1, ..., I_N, B_D\right]$ containing the states of the source, interior nodes, and destination for each active packet. In order to reduce the number of states, it is possible to group several \ac{CBR} states into a single Markov state.  For instance, the \ac{CBR} states corresponding to successful decoding and failure to decode a given packet by the destination can be grouped into two absorbing Markov state. The probability that the process moves from (Markov) state $s_i$ to state $s_j$ is denoted by $p_{i,j}$ and the probabilities $\{p_{i,j} \}$ are called {\em transition probabilities}.   Fig. \ref{Markov_Chain} shows the Markov chain that described the complete dynamics when there is only a single packet that propagates within a \ac{CBR} with $N=2$ relays (a four-node network).  The \ac{CBR} states are shown as a 4-element vector, along with the transition probabilities.  The $n^{th}$ message is successfully delivered whenever the destination receives the message, which is indicated by a $1_n$ in the last position of the \ac{CBR}-state vector.  Such states are marked in green, and hereafter we refer to this condition as a {\em \ac{CBR} success}.  The transmission fails whenever all entries in the \ac{CBR}-state vector, are either $0_n$ or $2_n$, indicating that none of the nodes that have not yet transmitted have successfully received the $n^{th}$ message.  Such states are marked in red, and hereafter we refer to this condition as a {\em \ac{CBR} outage}. In Fig. \ref{Markov_Chain} each transient state in the Markov process, which models the dynamics of the $n^{th}$ packet, are numbered 1 through 6 in the diagram. The 4 \ac{CBR} states that correspond to a \ac{CBR} outage are collapsed into a single absorbing state of the Markov process (state 7), and similarly the 9 \ac{CBR} states corresponding to a \ac{CBR} success are collapsed into a single absorbing state of the Markov process (state 8).

Consider how the \ac{CBR} state may change for the $n^{th}$ packet from time $t$ to time $t+1$.  All nodes at time $t$ with node state $1_n$ transmit the $n^{th}$ packet, so at time $t+1$, those nodes will now be in state $2_n$.  All nodes that have node state $0_n$ at time $t$ will receive the $n^{th}$ packet, so at time $t+1$ these nodes will be either in state $1_n$ (if the transmission was successful) or $0_n$ (if the transmission failed).  Since the channels from transmitting to receiving nodes are independent, the state transition probability is the product of the individual transmission probabilities.  For instance, the probability of going from state $[0_n 1_n 1_n 0_n ]$ to $[2_n 2_n 2_n 1_n]$ is the product of the probability that $B_D$ successfully decodes the joint transmission of the $n^{th}$ packet  from $I_1$ and $I_2$.  The individual probability of successful decoding at each node is found using the outage probability of (\ref{eqn_final_case1_Naka2}).  While there is a one-to-one correspondence between \ac{CBR} states and the transient Markov states, there are several \ac{CBR} states associated with each of the two absorbing Markov states.  Thus, for each absorbing Markov state, the probability of transitioning into it is equal to the sum of the probabilities of transitioning into the constituent \ac{CBR} states.

The Markov transition probabilities are placed into a \emph{state transition matrix} $\boldsymbol{P}^{(n)}$ for each $n^{th}$ packet, whose $(i,j)^{th}$ entry is $\{p_{i,j}\}$. The Markov states are ordered such that the $\mathcal{\tau}$ transient states and indexed before the $r$ absorbing states.  With this ordering, the state transition matrix assumes the following canonical form \cite{Grinstead:1997}
\begin{eqnarray}
\boldsymbol{P}^{(n)}= \left[\begin{matrix}
  \boldsymbol{Q}^{(n)} & \boldsymbol{R}^{(n)} \\
  \boldsymbol{0} & \boldsymbol{I}
 \end{matrix}\right]
\end{eqnarray}
where $\boldsymbol{Q}^{(n)}$ and $\boldsymbol{R}^{(n)}$ are respectively the $\tau \times \tau$ {\em transient matrix} and the $\tau \times r$ {\em absorbing matrix} for the $n^{th}$ packet, and $\boldsymbol{I}$ is an $r \times r$ identity matrix.

Let $b_{i,j}$ be the probability that the process will be absorbed in the absorbing state $s_j$ if it starts in the transient state $s_i$. The {\em absorbing probability} $b_{i,j}$ is the $(i,j)^{th}$ entry of the matrix $\boldsymbol{B}^{(n)}$, where as above the superscript $n$ indicates the packet under examination, which can be computed by (see, for instance \cite{Grinstead:1997})
\begin{eqnarray}
\boldsymbol{B}^{(n)}=  \boldsymbol{N}^{(n)} \boldsymbol{R}^{(n)} \label{absorbing_probabilities}
\end{eqnarray}
where $\boldsymbol{N}^{(n)}=\left(\boldsymbol{I}-\boldsymbol{Q}^{(n)} \right)^{-1}$ is the {\em fundamental matrix} for the $n^{th}$ packet. The two absorbing states are indexed so that the first absorbing state corresponds to a \ac{CBR} outage, while the second corresponds to a \ac{CBR}  success. Since the process always starts in state $s_1$, it follows that $b_{1,1}$ is the {\em \ac{CBR} outage probability} (which is indicated by $\epsilon_{CBR}$) and $b_{1,2}$ is the {\em \ac{CBR} success probability} (which is  $\hat{\epsilon}_{CBR}= 1- \epsilon_{CBR}$).

\section{Interference and Viterbi-Like Algorithm}
\label{Section:IterativeMethod}

 When the radio frame is sufficiently long to ensure that a packet reaches the destination before a new packet is injected within a \ac{CBR}, no collision occurs between packets. However, in practice \acp{BRN} often employ a fixed frame length, such as $F = 4$, and a fixed maximum number of relays that is greater than $F$  \cite{halford:2010a}. In such networks, spatial pipelining enables multiple packets, which will share the same radio resources, to be active simultaneously within a given \ac{CBR}. As previously mentioned, a new packet is injected in a given \ac{CBR} during the first time slot of each new radio frame, and when multiple packets propagate with a given \ac{CBR}, they interfere with each other. In order to account for this source of interference within the aforementioned framework and in order to account for the effect of correlation among packets an iterative method is describe in this section, which is based on a {\em Viterbi-like} algorithm.

\subsection{Iterative Algorithm}
\label{Section:IterativeMethod2}
Let $\mathbf P_k^{(n)}$ represent the transition matrix of the $k^{th}$ \ac{CBR} for the $n^{th}$ packet. When the inter-\ac{CBR} interference is neglected, but
multiple packets can simultaneously propagate within the same \ac{CBR}, there is a linkage between the transition probabilities for the $n^{th}$ packet $\mathbf P_k^{(n)}$ and the transmission probabilities of the nodes within the $k^{th}$ \ac{CBR} that may be active in letting other packets propagate during the interval of time the $n^{th}$ packet is broadcast. Furthermore, the probabilities of transmission for the $n^{th}$ packet  during a given radio frame depends also on the reciprocal interference produced by packets that have been propagating or had propagated within the $k^{th}$ \ac{CBR} since the very first transmission. These linkages could be handled by considering a large Markov chain, which describes the full dynamics in time of the $k^{th}$ \ac{CBR}. However, this solution is in most cases computational infeasible. A more appropriate approach is to use a Viterbi-like algorithm to iteratively update the transition matrix of the $k^{th}$ \ac{CBR} on a radio frame basis, alternating between the computation of the $\{\mathbf P_k^{(n)}\}$ and the $\{p_{i,n}^{(t)}\}$, where $p_{i,n}^{(t)}$ is the probability that $i^{th}$ node transmits the $n^{th}$ packet during time slot $t$.

Let $\boldsymbol{\mathcal N}[f]$ indicate the set of packets that might be active during the $f^{th}$ radio frame, which are ordered based on when they are first injected. Let $\boldsymbol{\mathcal N}[f](i)$ be the $i^{th}$ element of the set $\boldsymbol{\mathcal N}[f]$ and let $N_{pk}[f]=\big|\big|\boldsymbol{\mathcal N}[f]\big|\big|$ is their number, which is equal to $\displaystyle N_{pk}[f]=\left \lceil \frac{N+1}{F}\right \rceil $. Let $\boldsymbol{P}_{k}^{(n)}[i_t]$ represent the state transition matrix for the $k^{th}$ CBR at iteration $i_t$ for the $n^{th}$ packet, when it is active during the $f^{th}$ radio frame, and similarly let $p_{i,n}^{(t)}[i_t]$ represent the transmission probability at iteration $i_t$ for the $i^{th}$ user when it transmits the $n^{th}$ packet during the $t^{th}$ time slot of the $f^{th}$ radio frame. Let $\mathcal I_f$ indicate the maximum number of iterations needed to evaluate the transition matrix for the packet that is first transmitted during the $f^{th}$ radio frame. The iterative method can be described as follows:
\begin{enumerate}
\item Initialization: Set $f=1$ and $i_t=0$. Find the set $\boldsymbol{\mathcal N}[1]$ of packets that might be active during the first radio frame. Initialize ${p^{(t)}_{i,n}}[0]=0, \forall i,t$ and $n \in \boldsymbol{\mathcal N}[1]$. Note, that this corresponds to neglecting interference within the $k^{th}$ \ac{CBR}, during the duration of the first radio frame. Furthermore, it allows to start the iterative method from the scenario where the first packet injected within the $k^{th}$ \ac{CBR} is not subject to any intra-\ac{CBR} interference, unless it is still propagating during one of the successive radio frames. Compute $\boldsymbol{P}^{(n)}_{k}[0]$ using the initial $\{ p^{(t)}_{i,n}[0] \}$ for all $n \in \boldsymbol{\mathcal N}[1]$.
\item Recursion:
\begin{enumerate}
\item  Increment $i_t$. Compute $\boldsymbol{P}^{(n)}_{k}[i_t]$ using the $\{ p^{(t)}_{i,n}[i_t] \}$ from the previous iteration for all $n \in \boldsymbol{\mathcal N}[f]$. \label{stepIntra1}
\item Go back to step \ref{stepIntra1} until $||\left(\boldsymbol{P}_{k}^{(n)}[i_t]-\boldsymbol{P}_{k}^{(n)}[i_t-1]\right)||_F<\xi, \forall n \in \boldsymbol{\mathcal N}[f]$, where the operator $||\cdot||_F$ is the Frobenius norm and $\xi$ is a tolerance.
\end{enumerate}

\item Decision:
Halt the process if both the following conditions are met:
\begin{itemize}
\item $\Big|\Big| \left(\boldsymbol{P}_{k}^{\boldsymbol{\mathcal N}[f](i)}[\mathcal I_f ]-\boldsymbol{P}_{k}^{\boldsymbol{\mathcal N}[f-1](i)}[\mathcal I_{f-1} ]\right) \Big|\Big|_F  <  \xi$,  $\forall i\in \{1, \cdots, N_{pk}[f-1] \}$;
\item $f\neq1 $.
\end{itemize}
    Otherwise,
    \begin{enumerate}
\item   Increment $f$. Set $i_t=0$. Evaluate the set $\boldsymbol{\mathcal N}[f]$ of packets that might be active during the $f^{th}$ radio frame. \label{stepIntra3}
\item go back to step 2.
\end{enumerate}
\end{enumerate}

Note that through this approach it is possible to account not only for the reciprocal
interference that different packets produce to each other, but also for the effect of the temporal correlation among them.

\section{Network Optimization}
\label{Section:Optimization}

The \acl{TC} \cite{Gupta2000} of an ad hoc network quantifies the bits per second that can be reliably communicated over some distance in the network, and is typically expressed in units of meter-bits-per-second. In this context, the \ac{TC} of a typical \ac{CBR} is
\begin{eqnarray}
{\mathcal A}=  d_\text{CBR} \frac{(1-{\epsilon}_{CBR})}{ F} R
\label{tau}
\end{eqnarray}
where $R$ is the code rate, which is related here to the \ac{SINR} threshold by $R=\log_2\left( 1+\beta \right)$, which is the Shannon capacity for complex discrete-time AWGN channels.

As it is possible to notice from (\ref{tau}), there is a tradeoff between the code rate, the number of relays that compose a \ac{CBR}, and the number of time slots in a radio frame. Once the number of relays is fixed, as the number of time slots that compose the radio frame is reduced, the transport capacity tends to increase. Although the likelihood that multiple packets will be simultaneously active and collide within a \ac{CBR} is increased, with a consequent increment of the outage probability, which leads to a lower supported rate, and a reduction of the transport capacity.
When instead the number of relays is raised, the likelihood of multiple packets collision is increased, while on the other hand the network becomes more dense and links are shorter and might be more reliable. Therefore, the goal of the network optimization is to determine the set $\boldsymbol{\Theta}=\{R, N, F\}$ that maximizes (\ref{tau}).

In the following, assume that the inter-\ac{CBR} interference can be neglected. A line network is considered, which is composed of a \ac{CBR} with nodes disposed at equal distance from each other. Furthermore, assume that each node transmits with equal power $P_i=P$ and that the channels are Rayleigh faded.


Since the optimization surface is convex over $\{R, N, F\}$, while an exhaustive search could be a good alternative to this type of non-linear optimization problem, a more efficient approach to find the optimal set $\boldsymbol{\Theta}$ is as follows:
\begin{enumerate}
 \item For each of the parameters $N$, $R$ and $F$, select a pair of endpoints and a midpoint ($3^3 = 27$ points).  The endpoints should initially be far enough apart to guarantee that the optimal point lies within the endpoints.
 \item Compute the \ac{TC} by (\ref{tau}) for the endpoints and the midpoints of the interval of one of the parameters in the set $\{R, N, F\}$ by performing the iterative method described in Sec. \ref{Section:IterativeMethod2}. \label{step1a}
 \item For the same parameter of the set $\{R, N, F \}$ used in step \ref{step1a}, move the midpoint of the interval towards the endpoint that gives higher \ac{TC}. \label{step2a}
 \item Repeat step \ref{step1a} and \ref{step2a} recursively for all the parameters in the set $\{R, N, F\}$ and gradually reduce for one parameter at the time the range between the endpoints and the midpoint until a certain tolerance is reached. If the optimal \ac{TC} remains the same as in the previous iteration, the algorithm stops.
 \end{enumerate}

Using the convex optimization method described above, several optimizations are performed under different scenarios. Table \ref{Cap9_main_table2} shows the optimal values of $N$, $R$, and $F$, under all the scenarios under investigation, and provides for all of them the optimal \ac{TC} achieved. Optimization results are provided in Table \ref{Cap9_main_table2} for three values of $\Gamma$ and for each value of the \ac{SNR} three lengths for the \ac{CBR} are considered. For all these scenarios, it is assumed that the path loss is fixed to $\alpha=3.5$. Table \ref{Cap9_main_table2} shows that as expected the optimal \ac{TC} increases by increasing  the \ac{SNR}. More importantly, it is emphasized that as  the \ac{SNR} increases, it is possible to inject more frequently new packets by reducing the length of the radio frame and slightly the code rate. As the length of the \ac{CBR} is reduced, a fewer number of relays is required, until a point-to-point communication between the source and the destination is supported. Furthermore, while as expected the throughput increases as the length of the \ac{CBR} decreases, a lower \ac{TC} is achieved.

\begin{table}[!tbp]
\caption{Optimization results.}
\centering
  \begin{tabular}{|c||c||c||c||c||c|}
  \hline
  $d_\text{CBR}$ & $\mathsf{SNR}$ (in dB) & $R$ & $N$ & $F$ & ${\mathcal A}_\text{opt}$\\
  \hline
  \hline
  1 & 0        &     4.389    &   7   &  3  & 1.361  \\
  \cline{2-6}
  & 5     &     5.676   &   7   &  3 &  1.811  \\
  \cline{2-6}
   & 10       &    5.028    &   6   &  2 &  2.509  \\
  \hline
  \hline
  0.5 & 0        &     5.028    &   5   &  2  & 1.256  \\
  \cline{2-6}
  & 5     &     6.658   &   5   &  2 &  1.662  \\
  \cline{2-6}
   & 10       &    8.309    &   5   &  2 &  2.062  \\
 \hline
 \hline
  0.25 & 0        &     8.309    &   5   &  2  & 1.036  \\
  \cline{2-6}
  & 5     &     6.329   &   0   &  1 &  1.301  \\
  \cline{2-6}
   & 10       &    7.978    &   0   &  1 &  1.639  \\
   \hline
  \end{tabular}
    \label{Cap9_main_table2}
    \vspace{-0.35cm}
\end{table}

\section{Conclusion} \label{Section:Conclusion}

This paper has presented a novel analytical framework to model and more efficiently analyze \acp{BRN} that employ \acp{CBR} for unicast transmission. The performance of a \ac{BRN} is evaluated by describing the behavior of each constituent \acs{CBR} as a Markov chain, and
the transition probabilities are computed in closed form using a novel expression for the outage probability, which accounts for Rayleigh fading, co-channel interference, and path loss. The outage probability is conditioned over a given topology, but it is averaged over the fading allowing considerable reduction in the computational effort required to evaluate the averaged performance of the network.

The average end-to-end outage probability is evaluated in a direct manner, and does not involve the conditioning over a set of fading gains followed by the averaging over the fading and interference.   The interference caused by collisions between multiple packets that might be simultaneously active during a radio frame are taken into account. The temporal correlation as well as the interdependence of interfering transmissions is effectively captured and taken into account using a Viterbi-like algorithm, which iteratively updates the transition probabilities on frame-by-frame basis. Thanks to the reduction in computational effort, the iteratively evaluated analysis can be used to empower an optimization of relatively large \acp{CBR} for several different scenarios with the objective of maximizing the \ac{TC} with respect to the length of a radio frame, the code rate at which a node transmits and the number of relays involved in the communication. While the analysis and the model have been applied to study and optimize a \ac{BRN}, this work could be applied and extended to different other types of cooperative ad hoc networks that perform diversity combining of multiple source transmissions, for example, opportunistic large arrays, relay-selection systems, cooperative ARQ, and distributed space-time codes.

\balance

\bibliographystyle{ieeetr}
\bibliography{Milcom16ref}

\end{document}